\documentclass[
  reprint,
  amsmath,amssymb,
  aps,
  pra,
  superscriptaddress,
  floatfix,
]{revtex4-2}

\usepackage{graphicx}
\usepackage{dcolumn}
\usepackage{bm}
\usepackage[colorlinks=true,linkcolor=blue,citecolor=blue,urlcolor=blue]{hyperref}
\usepackage{xcolor}
\usepackage{subcaption}

\begin{document}

\preprint{APS/123-QED}

\title{Heterogeneous quantum error-correcting codes}

\author{Omid Khosravani}
\thanks{These authors contributed equally.}
\affiliation{Duke Quantum Center, Duke University, Durham, North Carolina 27701, USA}
\affiliation{Department of Physics, Duke University, Durham, North Carolina 27708, USA}
\affiliation{Department of Electrical and Computer Engineering, Duke University, Durham, North Carolina 27708, USA}
\affiliation{Institute for Robust Quantum Simulation, University of Maryland, College Park, Maryland 20742, USA}

\author{Guillermo Escobar-Arrieta}
\thanks{These authors contributed equally.}
\affiliation{Escuela de F\'isica, Universidad de Costa Rica, San Jos\'e 2060, Costa Rica}
\affiliation{Escuela de Ingenier\'ia El\'ectrica, Universidad de Costa Rica, San Jos\'e 2060, Costa Rica}

\author{Kenneth R. Brown}
\affiliation{Duke Quantum Center, Duke University, Durham, North Carolina 27701, USA}
\affiliation{Department of Physics, Duke University, Durham, North Carolina 27708, USA}
\affiliation{Department of Electrical and Computer Engineering, Duke University, Durham, North Carolina 27708, USA}
\affiliation{Department of Chemistry, Duke University, Durham, North Carolina 27708, USA}
\affiliation{Institute for Robust Quantum Simulation, University of Maryland, College Park, Maryland 20742, USA}

\author{Mauricio Guti\'errez}
\affiliation{Escuela de Qu\'imica, Universidad de Costa Rica, San Jos\'e 2060, Costa Rica}

\date{\today}

\begin{abstract}
We introduce heterogeneous quantum error-correcting codes composed of qubit types with distinct error channels and study their performance in the code-capacity regime using maximum-likelihood tensor network decoding.
In the regime where both qubit types share the same noise bias but differ in physical error rate, placing noisier qubits in the bulk---where each error triggers more syndrome bits---and cleaner qubits on the boundary yields thresholds exceeding 0.4 (compared to $\sim\!0.2$ for the reverse placement) and improvements exceeding three orders of magnitude in logical error rate at high bias, with the advantage growing exponentially with code distance.
In the regime where both types share the same error rate but differ in bias, the optimal strategy reverses: placing high-bias (more predictable) qubits on the boundary increases the threshold from $0.292(5)$ to $0.360(9)$ at $\eta_{\text{high}} = 100$, and from $0.29(1)$ to $0.398(4)$ at $\eta_{\text{high}} = 1000$.
We also observe a striking bias-inversion property: the logical error channel becomes strongly $X$- and $Y$-biased despite the physical noise being $Z$-biased.
We propose a stabilizer-ratio hypothesis that provides a unified information-theoretic explanation for both placement rules and predicts even larger advantages for code families such as color codes.
\end{abstract}

\maketitle

\section{Introduction}
\label{sec:intro}

The threshold theorem for fault-tolerant quantum computation~\cite{AharonovBenOr1997} establishes that arbitrarily long computations can be performed reliably provided that the physical error rate falls below a critical value, as recently demonstrated experimentally~\cite{AcharyaNature2023}. Historically, theoretical frameworks have predominantly considered homogeneous qubit systems with uniform error models, particularly the Pauli depolarizing channel~\cite{Fowler2012, Terhal2015}. This simplified model, while mathematically tractable, fails to capture the complexity of real quantum systems where different qubits exhibit varying error magnitudes, error channels may be strongly biased toward specific error types, and physical implementations naturally lead to distinct error characteristics for different qubit types.

These considerations are not merely theoretical: heterogeneous qubit systems are an emerging reality across multiple quantum computing platforms. In trapped-ion processors, different ionic species can serve as data and ancilla qubits with vastly different coherence times and gate fidelities~\cite{BruzewiczAPR2019}. Bosonic systems such as cat qubits exhibit strongly biased noise with bias ratios $\eta$ that can exceed 100~\cite{RegladeCatQubit2024,Puri2020}, while other qubits on the same chip may have near-depolarizing noise. Superconducting architectures increasingly integrate transmon qubits of varying quality, where fabrication variation alone produces a spread in $T_1$ and $T_2$ times across the device~\cite{Klimov2018}. Rather than treating this heterogeneity as an obstacle to be mitigated, a natural question arises: can we design error-correcting codes and decoding strategies that actively exploit the differences between qubit types to achieve better logical error rates?

Single-qubit Clifford deformations have proven to be a useful tool for improving the performance of codes under biased noise~\cite{TuckettPRL2018,BonillaNatComm2021,TiurevQuantum2023,ClaesNPJQI2023,DuaPRXQ2024}. The XY surface code~\cite{tuckett2018tailoring} replaces the $Z$ stabilizers of the standard surface code with $Y$ stabilizers so that $Z$ errors anticommute with both the $X$ and $Y$ stabilizer types, providing more syndrome information about the dominant error. In the code-capacity regime with a tensor network (TN) decoder, the XY surface code achieves thresholds that increase monotonically with the noise bias, from $18.7\%$ for the depolarizing channel to $43.7\%$ for pure dephasing~\cite{tuckett2018tailoring}. The XZZX surface code~\cite{BonillaNatComm2021} achieves high thresholds even with a simple matching decoder. As shown in Ref.~\cite{TiurevQuantum2023}, it is possible to tailor surface codes to heterogeneous single-qubit noise by carefully aligning the Pauli noise and the stabilizer checks at each site.

Crucially, these ideas extend well beyond the surface code. The domain wall color code~\cite{TiurevDWCC2024} achieves a 50\% code-capacity threshold at infinite bias and matches the XZZX surface code thresholds at finite bias, while retaining the color code's transversal Clifford gate advantage. The XYZ color code~\cite{SanMiguelXYZ2023} exhibits fracton-like properties under infinite dephasing bias, eliminating string-like logical operators entirely. Clifford deformations have been applied to three-dimensional topological codes including the 3D color code~\cite{HuangPRXQ2023}, to compass codes~\cite{CamposCompass2024}, and to quantum LDPC codes~\cite{RoffeLDPC2023}. However, all of these studies consider homogeneous noise across the code---no prior work has examined the strategic placement of distinct qubit types within any of these bias-tailored code families. The concept of heterogeneous quantum error-correcting codes, in which qubits of fundamentally different types are combined within a single code block, was introduced in Ref.~\cite{KhosravaniThesis2024} and presented in Ref.~\cite{KhosravaniAPS2025}. We note that the term ``heterogeneous'' has been used in recent QEC literature to describe architectures that combine different code families within a single fault-tolerant system~\cite{SteinHetEC2024,SteinbergHolographic2025}; our usage is distinct, referring to physically different qubit types---with different error rates or noise biases---strategically placed within a single code block.

In this work, we study the problem of qubit placement within Clifford-deformed surface codes for heterogeneous systems with two types of qubits. We investigate two regimes: (a)~qubits with the same bias but different physical error rates, and (b)~qubits with the same physical error rate but different bias values. Using tensor network decoding, we find that strategic placement of qubits can yield improvements exceeding three orders of magnitude in logical error rate and substantial threshold gains. We observe a striking bias-inversion property in the logical error channel and propose a stabilizer-ratio hypothesis that explains both placement rules and predicts even larger advantages for other code families.

\section{Setup}
\label{sec:setup}

\subsection{Biased noise model}

The biased noise error channel is modeled as
\begin{equation}
\label{eq:biased}
\begin{split}
\mathcal{E}_{\text{biased}}(\rho) &= (1-p)\rho \\
&\quad + p\!\left(\frac{\eta}{1+\eta} Z\rho Z + \frac{X\rho X + Y\rho Y}{2(1+\eta)}\right)\!,
\end{split}
\end{equation}
where $p$ is the total error probability and $\eta = p_Z / (p_X + p_Y)$ is the bias ratio, following the convention of Ref.~\cite{TuckettPRL2018}. This gives $p_Z = p\eta/(1+\eta)$ and $p_X = p_Y = p/[2(1+\eta)]$. When $\eta = 1/2$, this reduces to the standard depolarizing channel; as $\eta \to \infty$, the channel approaches pure dephasing.

\subsection{Clifford deformations and the XY surface code}

Clifford deformations consist of conjugating the stabilizers of a QEC code by single-qubit Clifford operators, transforming a code into a topologically equivalent one with vastly different performance against biased noise. Among the Clifford-deformed versions of the rotated surface code catalogued in Ref.~\cite{DuaPRXQ2024}, the XY code does not have the highest threshold at all bias values---at the moderate biases ($\eta = 10$--$100$) considered here, other deformations such as C1 and C2 can outperform it in the homogeneous setting. We nevertheless focus on the XY surface code because it applies the \emph{same} single-qubit Clifford to every data qubit, making it the natural starting point for heterogeneous codes: the optimal Clifford deformation for a site with one error channel need not be optimal for a site with a different channel, and the uniform structure of the XY code avoids conflating the effects of heterogeneous placement with those of non-uniform deformations. Single-qubit deformations do not change the overall code distance $d$ but can change the individual Pauli distances $d_x$, $d_y$, and $d_z$, allowing the distance along the dominant error type to be increased.

\subsection{Tensor network decoding}

Maximum-likelihood decoding can be recast as a tensor network contraction~\cite{PoulinQECTN2014,Darmawan2017}. Following Bravyi, Suchara, and Vargo~\cite{Vargo2014}, a TN structure is set up based on the code's geometry, with tensors representing qubits and stabilizers. An approximate contraction with fixed bond dimension $\chi$ yields a practical decoder. We employ the Sweep Contractor implementation by Chubb~\cite{Chubb2D,ChubbGithub} with bond dimension $\chi = 16$ for all threshold estimates, which we verified is sufficient by checking select data points at $\chi = 48$ (threshold values shifted by less than 1\% in all cases tested). Error thresholds are estimated using the critical exponent method of Wang \emph{et al.}~\cite{Wang2003}; all reported threshold uncertainties are standard errors from this fitting procedure. We note that with only three code distances ($d = 5, 7, 9$), the finite-size scaling fits have limited degrees of freedom; the reported thresholds should be interpreted as estimates whose precision will improve with data at larger distances.

\subsection{Heterogeneous error model}

We consider a system with two types of qubits, each described by a biased noise channel of the form Eq.~\eqref{eq:biased}. We study two regimes chosen to isolate the effects of error rate heterogeneity and bias heterogeneity independently:

\paragraph*{Regime~A (different error rates, same bias).}
Both qubit types share the same bias $\eta$ but have total error rates differing by a factor of 10: $p_{\text{noisy}} = 10\, p_{\text{quiet}}$. We test $\eta = 10$ (moderate dephasing bias~\cite{AcharyaNature2023}) and $\eta = 100$ (strong bias, representative of cat qubits~\cite{RegladeCatQubit2024}). As a control, we also test $\eta = 0.5$ (the depolarizing channel, i.e.\ no bias) to verify that the placement advantage vanishes when the noise bias does not align with the Clifford deformation.

\paragraph*{Regime~B (same error rate, different biases).}
Both qubit types share the same total error rate $p$ but have different biases: $\eta_{\text{low}} = 10$ and $\eta_{\text{high}} \in \{100, 1000\}$, both favoring $Z$ errors. This models a hybrid processor combining mildly biased transmons with strongly biased cat qubits~\cite{Puri2020}.

In both regimes, the number of each qubit type is approximately equal, and we study three placement strategies (see Table~\ref{tab:configs} and Fig.~\ref{fig:surface_codes_xy}):

\begin{table}[b]
\centering
\caption{Placement strategies for the two heterogeneity regimes. ``Boundary'' refers to qubits touched by $\leq 3$ stabilizer generators; ``bulk'' refers to qubits touched by~4. Labels are defined by Regime~A semantics; in Regime~B, ``\textsc{Bulk-Noisy}'' denotes the configuration where the \emph{less predictable} (low-bias) qubits occupy the bulk.}
\label{tab:configs}
\begin{tabular}{lll}
\hline\hline
Label & Regime~A & Regime~B \\
\hline
\textsc{Bulk-Noisy} & noisy in bulk & low-bias in bulk \\
\textsc{Boundary-Noisy} & noisy on boundary & high-bias on boundary \\
\textsc{Random} & random placement & random placement \\
\hline\hline
\end{tabular}
\end{table}

All simulations are performed in the code-capacity regime with perfect syndrome measurements.

\section{Results}
\label{sec:results}

\subsection{Regime A: same bias, different physical error rates}

\begin{figure}[!htbp]
    \centering
    \begin{subfigure}[b]{0.145\textwidth}
        \includegraphics[width=\textwidth]{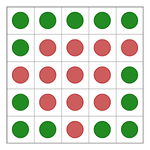}
        \caption{$d\!=\!5$}
    \end{subfigure}
    \hfill
    \begin{subfigure}[b]{0.145\textwidth}
        \includegraphics[width=\textwidth]{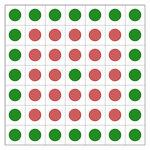}
        \caption{$d\!=\!7$}
    \end{subfigure}
    \hfill
    \begin{subfigure}[b]{0.145\textwidth}
        \includegraphics[width=\textwidth]{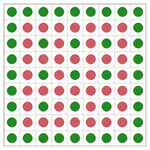}
        \caption{$d\!=\!9$}
    \end{subfigure}

    \vspace{0.2cm}

    \begin{subfigure}[b]{0.145\textwidth}
        \includegraphics[width=\textwidth]{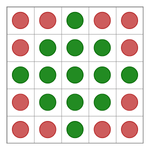}
        \caption{$d\!=\!5$}
    \end{subfigure}
    \hfill
    \begin{subfigure}[b]{0.145\textwidth}
        \includegraphics[width=\textwidth]{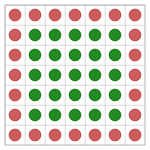}
        \caption{$d\!=\!7$}
    \end{subfigure}
    \hfill
    \begin{subfigure}[b]{0.145\textwidth}
        \includegraphics[width=\textwidth]{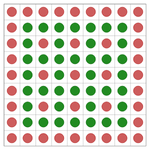}
        \caption{$d\!=\!9$}
    \end{subfigure}
    \caption{Rotated XY surface codes with \textsc{Bulk-Noisy} placement (top row, a--c) and \textsc{Boundary-Noisy} placement (bottom row, d--f) for distances 5, 7, and~9. Green (red) circles denote quiet/high-bias (noisy/low-bias) qubits. The number of each qubit type is approximately equal.}
    \label{fig:surface_codes_xy}
\end{figure}

We first study the case where both subsets of qubits share the same bias $\eta$ but differ in total error rate by a factor of 10 ($p_{\text{noisy}} = 10\, p_{\text{quiet}}$). Figure~\ref{fig:surface_codes_xy} shows the XY rotated surface codes with the \textsc{Bulk-Noisy} and \textsc{Boundary-Noisy} placements for distances 5, 7, and~9.

Table~\ref{table:thresholds_variable_p} summarizes the threshold estimates. For $\eta = 10$, \textsc{Bulk-Noisy} exhibits a threshold of at least 0.4 (in terms of $p_{\text{noisy}}$; the true threshold lies above our simulated range), while \textsc{Boundary-Noisy} achieves only $0.21(6)$---roughly half. For $\eta = 100$, the pattern persists: $0.43(2)$ versus $0.23(5)$. In both cases, \textsc{Bulk-Noisy} substantially outperforms \textsc{Boundary-Noisy}.

\begin{table}[tb]
\centering
\caption{Thresholds (in terms of $p_{\text{noisy}}$) for Regime~A: same bias $\eta$, error rate ratio $p_{\text{noisy}}/p_{\text{quiet}} = 10$. Parenthetical values are the standard error on the last digit. The entry ``$>0.4$'' indicates that the threshold lies above the simulated error-rate range and should be read as a lower bound.}
\label{table:thresholds_variable_p}
\begin{tabular}{ccc}
\hline\hline
$\eta$ & \textsc{Bulk-Noisy} & \textsc{Boundary-Noisy} \\
\hline
10  & $>0.4$   & $0.21(6)$ \\
100 & $0.43(2)$ & $0.23(5)$ \\
\hline\hline
\end{tabular}
\end{table}

\begin{figure}[!htbp]
    \centering
    \begin{subfigure}{0.23\textwidth}
        \centering
        \includegraphics[width=\textwidth]{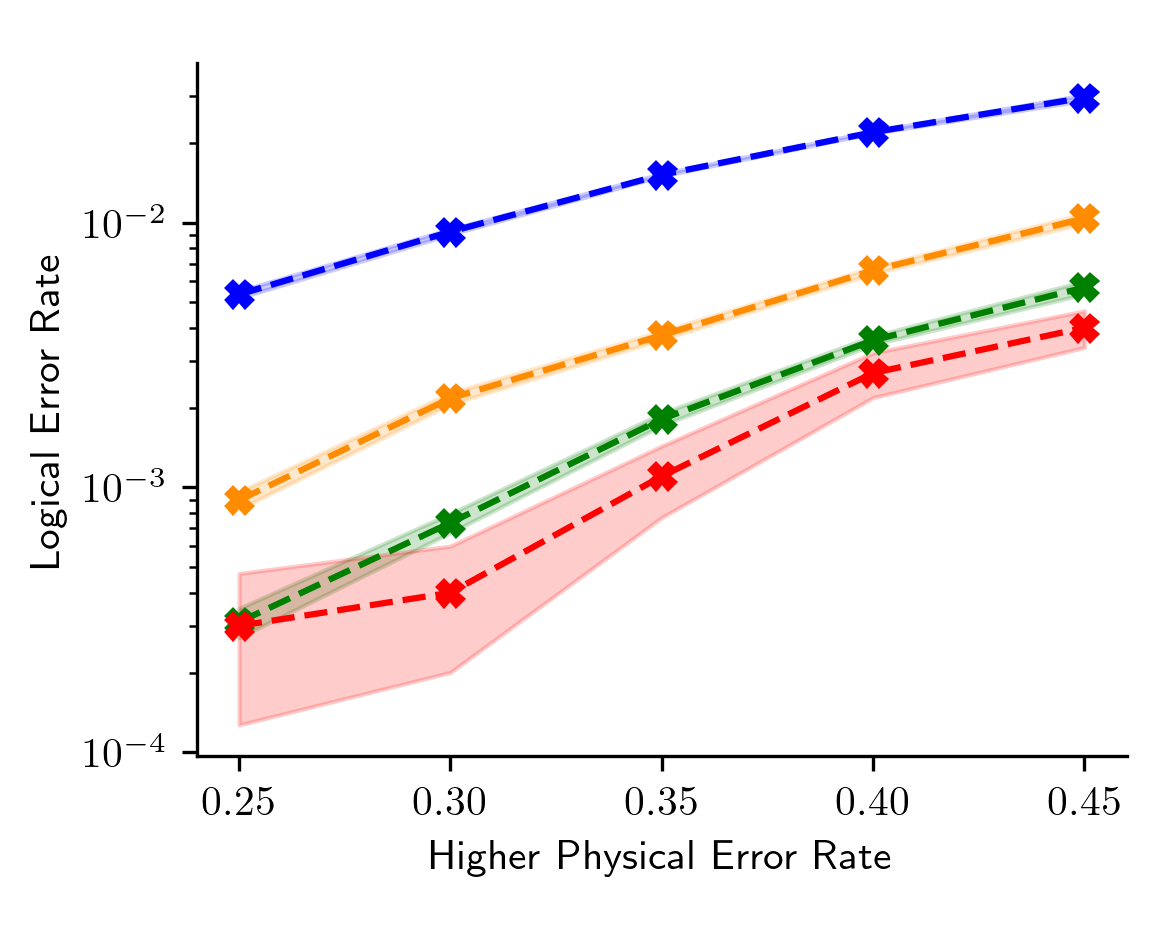}
        \caption{\textsc{Bulk-Noisy}}
    \end{subfigure}
    \begin{subfigure}{0.23\textwidth}
        \centering
        \includegraphics[width=\textwidth]{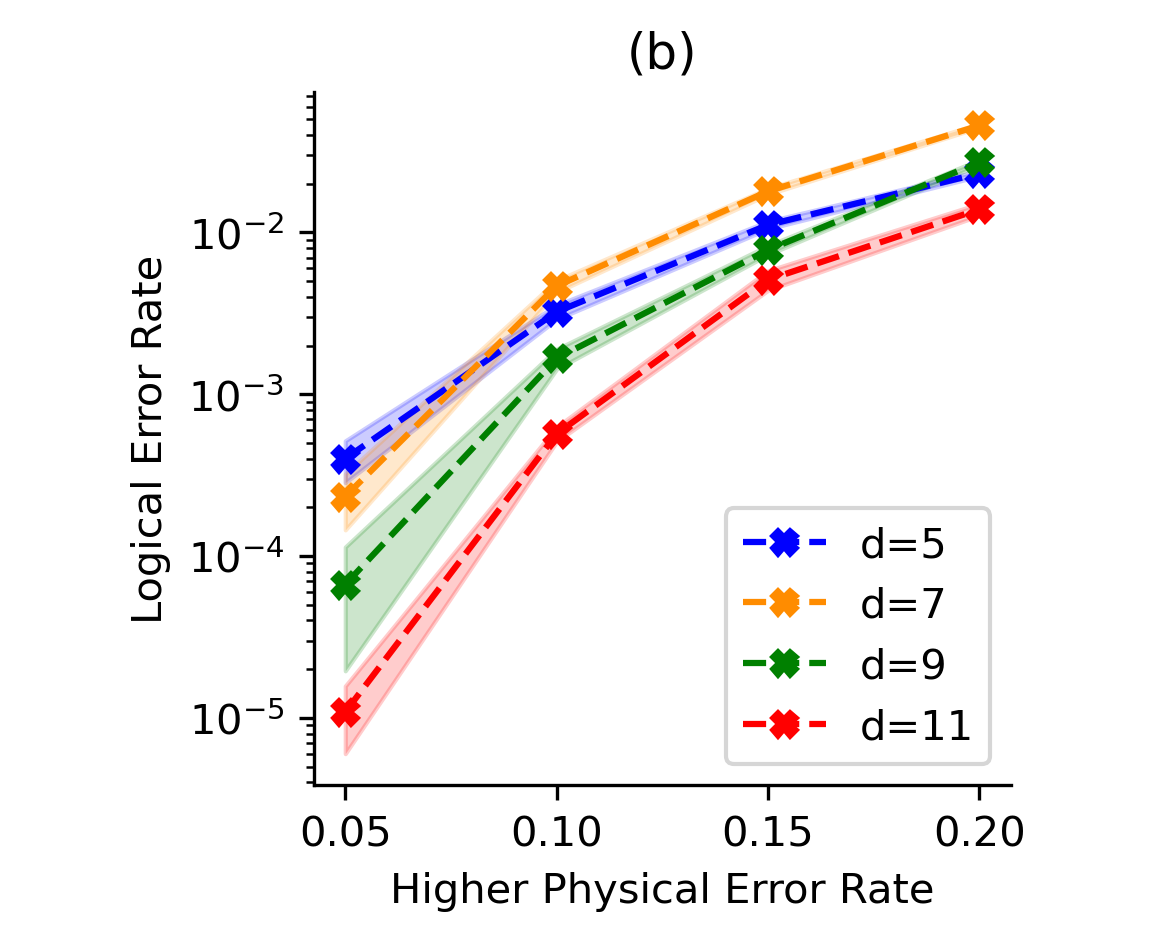}
        \caption{\textsc{Boundary-Noisy}}
    \end{subfigure}
    \caption{Logical failure rates for Regime~A with $\eta = 10$ and $p_{\text{noisy}}/p_{\text{quiet}} = 10$, for (a)~\textsc{Bulk-Noisy} (distances 5, 7, 9) and (b)~\textsc{Boundary-Noisy} (distances 5, 7, 9, 11). The crossing of the distance curves indicates the threshold.}
    \label{fig:variablep_eta10}
\end{figure}

\begin{figure}[!htbp]
    \centering
    \includegraphics[width=0.9\columnwidth]{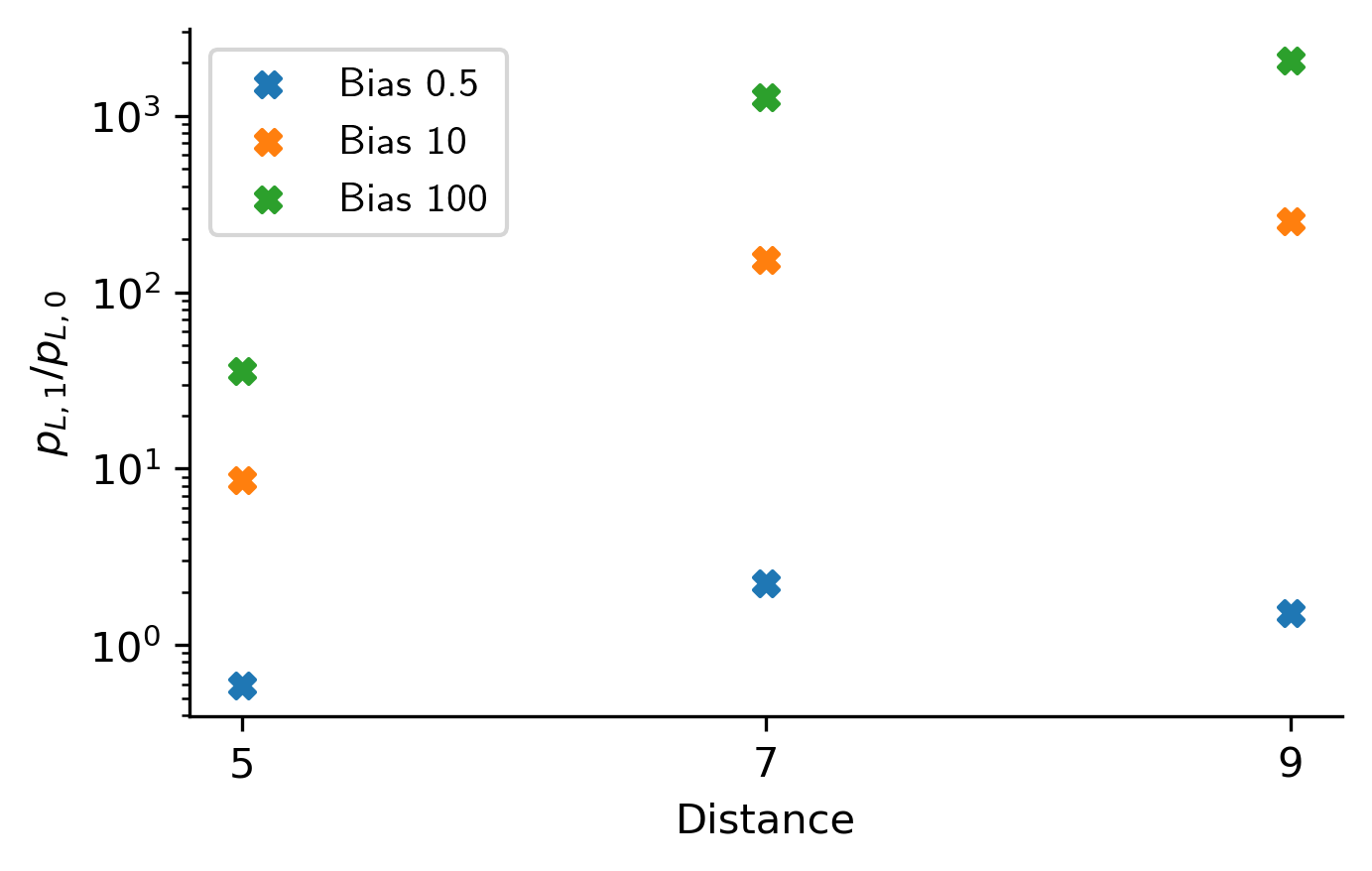}
    \caption{Ratio of logical error rates $p_{L,1}/p_{L,0}$ between \textsc{Boundary-Noisy} and \textsc{Bulk-Noisy} as a function of code distance, for $p_{\text{noisy}} = 0.30$ and $p_{\text{noisy}}/p_{\text{quiet}} = 10$, at three bias values. At $\eta = 100$ the advantage exceeds $10^3$ by distance~9. The $\eta = 0.5$ control (depolarizing channel) confirms that the placement effect requires alignment between the noise bias and the Clifford deformation.}
    \label{fig:improv_rate}
\end{figure}

Figure~\ref{fig:variablep_eta10} shows the logical failure rates for $\eta = 10$. The crossing points of the distance curves confirm the threshold estimates in Table~\ref{table:thresholds_variable_p}. Notably, the advantage of \textsc{Bulk-Noisy} over \textsc{Boundary-Noisy} grows rapidly with code distance: Fig.~\ref{fig:improv_rate} shows that for $p_{\text{noisy}} = 0.30$ and $p_{\text{noisy}}/p_{\text{quiet}} = 10$, the ratio of logical error rates grows exponentially with $d$, exceeding $10^3$ at $\eta = 100$ and $d = 9$. The growth rate increases with bias, and the $\eta = 0.5$ control (depolarizing channel) confirms that the placement advantage is negligible when the noise bias does not align with the XY deformation. This exponential growth at high bias implies that the two placements have effectively different thresholds, so the gap will continue to widen at larger distances.

\subsection{Regime B: same physical error rate, different biases}

Next, we study the regime where all qubits have the same physical error rate $p$ but differ in bias: $\eta_{\text{low}} = 10$ and $\eta_{\text{high}} \in \{100,\, 1000\}$. Note that the configuration labels from Table~\ref{tab:configs} carry over from Regime~A: \textsc{Bulk-Noisy} now means low-bias (less predictable) qubits in the bulk and high-bias qubits on the boundary, while \textsc{Boundary-Noisy} means the reverse.

A notable reversal occurs relative to Regime~A: \textsc{Bulk-Noisy} (high-bias on boundary) now \emph{outperforms} \textsc{Boundary-Noisy}, even though all qubits have the same total error rate. Table~\ref{table:thresholds_variable_eta} shows the results. For $\eta_{\text{high}} = 100$, \textsc{Bulk-Noisy} achieves $p_{\text{th}} = 0.360(9)$ compared to $0.292(5)$ for \textsc{Boundary-Noisy}---a 23\% increase. For $\eta_{\text{high}} = 1000$, the gap widens further: $0.398(4)$ versus $0.29(1)$---a 37\% increase, approaching the infinite-bias limit. The \textsc{Random} placement falls between the two in both cases.

\begin{table}[tb]
\centering
\caption{Thresholds for Regime~B: same physical error rate $p$, $\eta_{\text{low}} = 10$, varying $\eta_{\text{high}}$. Parenthetical values are the standard error on the last digit.}
\label{table:thresholds_variable_eta}
\begin{tabular}{cccc}
\hline\hline
$\eta_{\text{high}}$ & \textsc{Bulk-Noisy} & \textsc{Boundary-Noisy} & \textsc{Random} \\
\hline
100  & $0.360(9)$ & $0.292(5)$ & $0.321(5)$ \\
1000 & $0.398(4)$ & $0.29(1)$  & $0.33(2)$  \\
\hline\hline
\end{tabular}
\end{table}

\begin{figure}[!htbp]
    \centering
    \includegraphics[width=0.9\columnwidth]{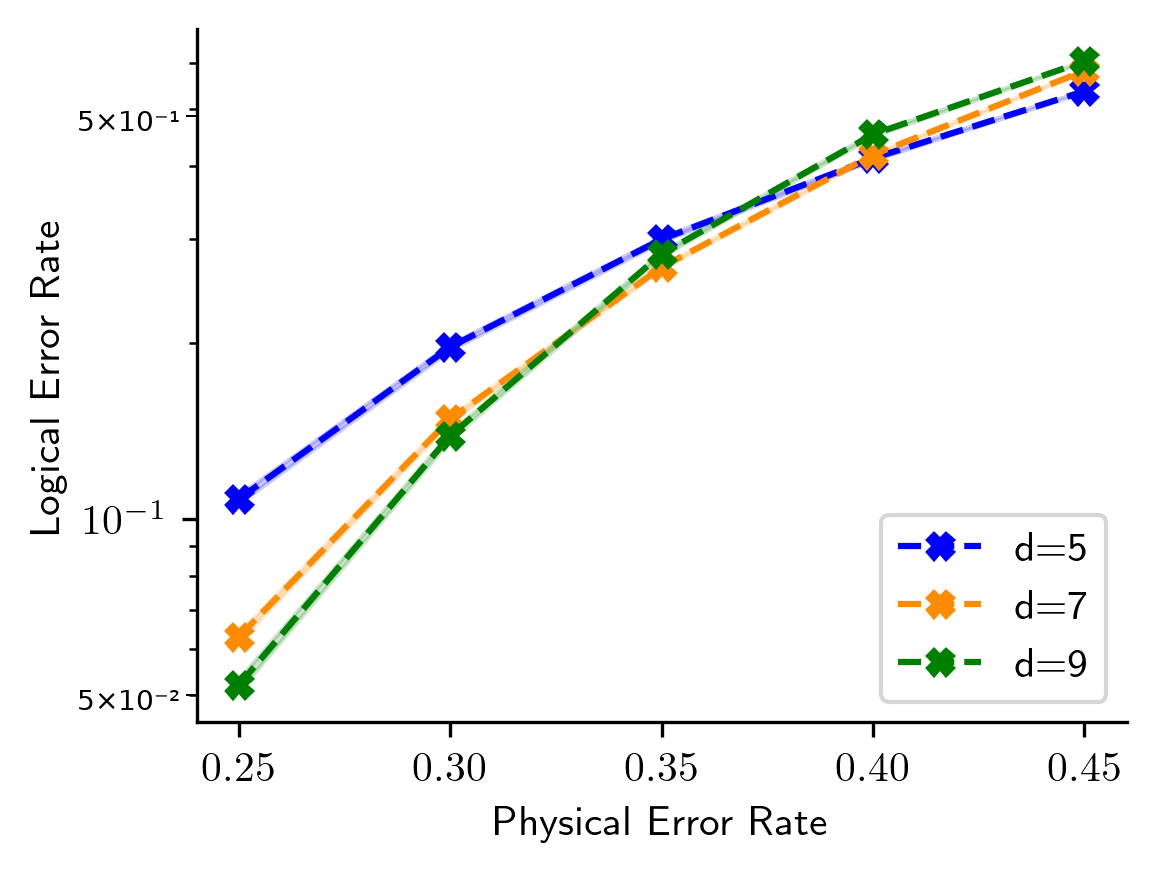}
    \caption{Logical failure rates for Regime~B with \textsc{Bulk-Noisy} placement, $\eta_{\text{low}} = 10$, $\eta_{\text{high}} = 100$, for distances 5, 7, and~9. The crossing near $p \approx 0.36$ confirms the threshold estimate in Table~\ref{table:thresholds_variable_eta}.}
    \label{fig:variableeta}
\end{figure}

Figure~\ref{fig:variableeta} shows the logical failure rates for \textsc{Bulk-Noisy} with $\eta_{\text{high}} = 100$. The crossing point near $p \approx 0.36$ is consistent with the threshold in Table~\ref{table:thresholds_variable_eta}.

\subsection{Logical channel properties: bias inversion}

\begin{figure}[!htbp]
    \centering
    \includegraphics[width=0.9\columnwidth]{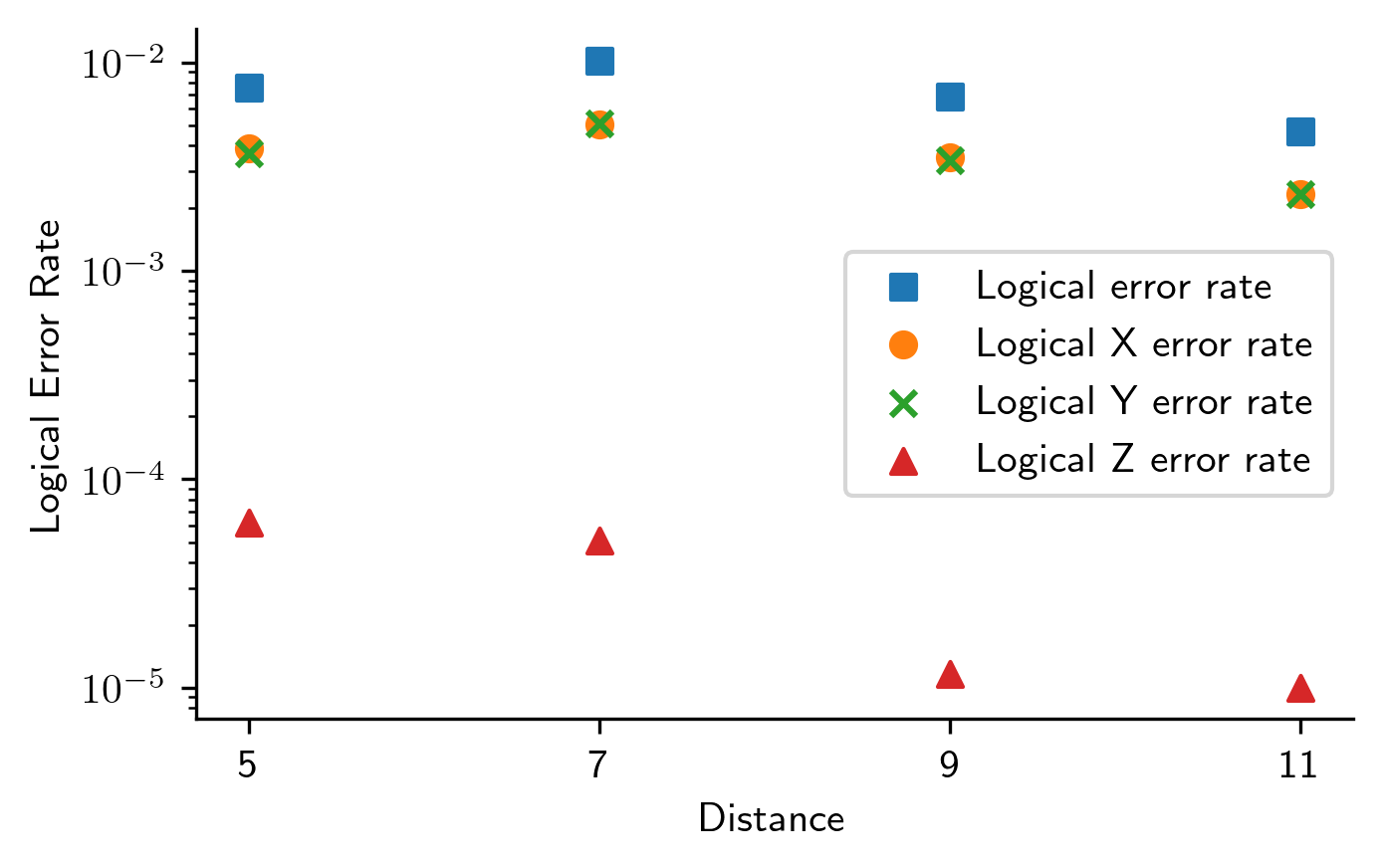}
    \caption{Disaggregated logical Pauli error rates for \textsc{Boundary-Noisy}, with $p_{\text{noisy}} = 0.2$, $p_{\text{noisy}}/p_{\text{quiet}} = 10$, $\eta = 100$, for distances 5--11. Despite the physical noise being strongly $Z$-biased, the logical channel is dominated by $X_L$ and $Y_L$ errors (which nearly overlap), with a logical bias $\eta_L \approx 4 \times 10^{-3}$. The $Z_L$ error rate is suppressed by over two orders of magnitude.}
    \label{fig:disaggregate}
\end{figure}

A striking finding is that the logical error channel exhibits \emph{bias inversion}. When using physical qubits with strongly $Z$-biased noise, the resulting logical channel becomes dominated by $X_L$ and $Y_L$ errors. Figure~\ref{fig:disaggregate} shows the disaggregated logical Pauli error rates for \textsc{Boundary-Noisy} at $p_{\text{noisy}} = 0.2$, $p_{\text{noisy}}/p_{\text{quiet}} = 10$, and $\eta = 100$. Most logical failures arise from $X_L$ and $Y_L$ errors, with approximately equal contributions, yielding a logical bias $\eta_L = p_{Z_L}/(p_{X_L} + p_{Y_L}) \approx 4 \times 10^{-3}$. In fact, the inversion strengthens with code distance: $Z_L$ errors are suppressed more rapidly than $X_L$ and $Y_L$ errors as $d$ grows, so $\eta_L$ decreases monotonically.

This bias-inversion phenomenon has been observed in homogeneous bias-tailored codes~\cite{TuckettPRL2018,tuckett2018tailoring}, but behaves distinctively in the heterogeneous setting. The inversion arises because the XY deformation increases $d_z$ relative to $d_x$ and $d_y$: the dominant physical $Z$ errors are efficiently suppressed, while the rare $X$ and $Y$ errors---which traverse the shorter logical distance---dominate the residual logical failure rate. More concretely, although the physical $X$ and $Y$ error rates on the noisy qubits are small (on the order of $0.002$--$0.004$, well below the single-channel threshold of $\sim\!0.1$ for the standard surface code), the XY code's deformation effectively places the $X/Y$ error channel close to its reduced threshold while the $Z$ channel is driven deep below threshold. This is the hallmark of a two-channel decoding regime: when one Pauli channel ($Z$) is far below its effective threshold but the other ($X/Y$) is only marginally below, the $Z_L$ component of the logical error rate drops rapidly with distance while $X_L$ and $Y_L$ decrease much more slowly, producing the observed bias inversion. The inversion strengthens with distance precisely because the gap between the two channels' proximity to their respective thresholds grows. Bias inversion also occurs in the \textsc{Bulk-Noisy} configuration (not shown) with comparable $\eta_L$, confirming that it is a property of the XY deformation under heterogeneous noise rather than an artifact of a particular placement.

The bias-inversion property has practical implications for concatenated error correction: the outer code protecting a heterogeneous inner code should be designed for $X/Y$-biased (rather than $Z$-biased) logical noise.

\section{Discussion: the stabilizer-ratio hypothesis}
\label{sec:discussion}

The two regimes yield placement rules that appear opposite but share a common information-theoretic origin.

In Regime~A (same bias, different error rates), the optimal strategy places noisier qubits in the bulk. Bulk qubits in the rotated 2D surface code are touched by 4 stabilizer generators, while boundary qubits are touched by only 2 (corners) or 3 (edges). When a bulk qubit suffers an error, it triggers up to 4 syndrome changes, providing more information to the decoder than a boundary error that triggers only 2 or~3. Placing the noisier---and hence more frequently erring---qubits in the bulk ensures that their errors are diagnosed with maximal syndrome information.

In Regime~B (same error rate, different biases), the optimal strategy reverses: high-bias qubits go on the boundary. A high-bias qubit's errors are already highly predictable---the decoder knows \emph{a priori} that nearly all errors will be $Z$ type. This prior predictability partially compensates for the reduced syndrome information at the boundary. Conversely, low-bias qubits produce less predictable errors and benefit more from the extra syndrome bits available in the bulk.

Both observations follow from a single principle: \emph{the qubits whose errors are hardest to decode should be placed where the decoder has the most information}. We formalize this by defining the stabilizer ratio
\begin{equation}
\label{eq:r}
r = \frac{\text{avg.\ stabilizers per bulk qubit}}{\text{avg.\ stabilizers per boundary qubit}}.
\end{equation}
For the rotated surface code, $r$ can be computed explicitly. A distance-$d$ rotated code has $d^2$ data qubits: $(d-2)^2$ bulk qubits (each touched by 4 stabilizers), $4(d-2)$ edge qubits (touched by 3), and 4 corner qubits (touched by 2). Thus the average stabilizer count per boundary qubit is $[3 \cdot 4(d-2) + 2 \cdot 4] / [4(d-2)+4] = (3d-2)/(d-1)$, and the bulk average is~4, giving
\begin{equation}
r(d) = \frac{4(d-1)}{3d-2}\,.
\end{equation}
This evaluates to $r(5) = 16/13 \approx 1.23$, $r(9) = 32/25 = 1.28$, and $r \to 4/3 \approx 1.33$ as $d \to \infty$. The information asymmetry between bulk and boundary grows with $r$: larger $r$ means the decoder's ability to diagnose errors depends more strongly on qubit position. Consequently, the advantage of optimal placement should grow with $r$.

Two caveats apply. First, the fraction of boundary qubits scales as $\sim\!4/d$ for a distance-$d$ rotated surface code, so at large $d$ the placement choice affects a diminishing fraction of qubits. The large advantages we observe at $d = 5$--$9$ partially reflect that boundary qubits constitute a significant minority at these distances. However, the exponential growth of the improvement ratio (Fig.~\ref{fig:improv_rate}) indicates that the effect remains consequential even as boundary fraction shrinks. Second, $r$ alone does not capture the full story: logical operator structure, decoder choice, and boundary geometry all play roles. We present the stabilizer-ratio hypothesis as a useful heuristic, rather than a precise quantitative law.

\paragraph*{Predictions for other code families.}
The 2D color code on the 6.6.6 (honeycomb) lattice has $r \approx 3/2 > 4/3$. Our hypothesis predicts that heterogeneous placement in color codes should yield even larger advantages. This is directly testable using the domain wall color code~\cite{TiurevDWCC2024}, which achieves a 50\% threshold at infinite bias while retaining transversal Clifford gates, or the XYZ color code~\cite{SanMiguelXYZ2023}, which eliminates string-like logical operators under infinite bias. In 3D, Clifford-deformed topological codes~\cite{HuangPRXQ2023} have even larger bulk-to-boundary ratios, and Clifford-deformed compass codes~\cite{CamposCompass2024} offer tunable gauge fixing that could be co-optimized with qubit placement.

\paragraph*{Relation to prior work and open questions.}
Our approach is complementary to that of Tiurev \emph{et al.}~\cite{TiurevQuantum2023}, who optimize the \emph{code} to match given noise, while we optimize the \emph{placement of qubit types} within a fixed code. In principle, both could be combined. Open questions include: how does heterogeneous placement interact with the XZZX deformation~\cite{BonillaNatComm2021} or with bias-tailored LDPC codes~\cite{RoffeLDPC2023}?

On a real device where qubit types differ in \emph{both} error rate and bias simultaneously, our two regimes yield opposite placement rules. Qualitatively, the error-rate asymmetry (Regime~A) enters through syndrome frequency (linear in the rate ratio), while the bias asymmetry (Regime~B) enters through prior predictability (logarithmic in the bias ratio). We therefore expect Regime~A to dominate when $p_{\text{noisy}}/p_{\text{quiet}} \gg \eta_{\text{high}}/\eta_{\text{low}}$, and vice versa. Mapping this crossover boundary is an important direction for future work.

\section{Conclusion}
\label{sec:conclusion}

We have introduced and studied heterogeneous quantum error-correcting codes in which distinct qubit types are strategically placed within a Clifford-deformed surface code. For qubits sharing the same bias but differing in error rate, placing noisier qubits in the syndrome-rich bulk yields thresholds of at least 0.4, compared to $\sim\!0.2$ for the reverse placement, with the logical error rate advantage growing exponentially with code distance. For qubits sharing the same error rate but differing in bias, placing high-bias qubits on the boundary yields up to a 37\% threshold increase (from $0.29$ to $0.40$ at $\eta_{\text{high}} = 1000$). We have identified a bias-inversion property of the logical channel and proposed a stabilizer-ratio hypothesis that unifies both placement rules and predicts even larger advantages for color codes ($r \approx 3/2$), 3D codes, and other topological code families.

All results are obtained in the code-capacity regime; extending to circuit-level noise is an important next step. Comparison with the XZZX code~\cite{BonillaNatComm2021}, direct testing on color codes~\cite{TiurevDWCC2024,SanMiguelXYZ2023}, and benchmarking against the homogeneous (single qubit type) baseline are natural follow-ups, as is a systematic crossover analysis for devices with simultaneous error-rate and bias heterogeneity. Our results suggest that future quantum processors should be designed with heterogeneous qubit placement in mind: by combining different physical qubit types and assigning them to code positions according to the stabilizer-ratio principle, one can achieve substantially lower logical error rates at the same code distance compared to naive or random placement.

The scripts used to generate the data presented in this work are available at \url{PLACEHOLDER_URL}.

\begin{acknowledgments}
This project was supported in part by the NSF Quantum Leap Challenge Institute (QLCI), the NSF Software-Tailored Architecture for Quantum co-design (STAQ) project, the National Energy Research Scientific Computing Center (NERSC), a DOE Office of Science User Facility supported by the Office of Science of the U.S. Department of Energy under Contract No. DE-AC02-05CH11231 using NERSC award ERCAP0025994, and the ORISE Postdoctoral Research Fellowship program, administered by Oak Ridge Institute for Science and Education (ORISE) under the U.S. Department of Energy.
\end{acknowledgments}

\bibliography{references}

@article{AharonovBenOr1997,
  author  = {Aharonov, Dorit and Ben-Or, Michael},
  title   = {Fault-tolerant quantum computation with constant error},
  journal = {Proceedings of the 29th Annual ACM Symposium on Theory of Computing},
  pages   = {176--188},
  year    = {1997},
  doi     = {10.1145/258533.258579}
}

@article{AcharyaNature2023,
  author  = {Acharya, Rajeev and others},
  title   = {Suppressing quantum errors by scaling a surface code logical qubit},
  journal = {Nature},
  volume  = {614},
  pages   = {676--681},
  year    = {2023},
  doi     = {10.1038/s41586-022-05434-1}
}

@article{BruzewiczAPR2019,
  author  = {Bruzewicz, Colin D. and Chiaverini, John and McConnell, Robert and Sage, Jeremy M.},
  title   = {Trapped-ion quantum computing: Progress and challenges},
  journal = {Applied Physics Reviews},
  volume  = {6},
  pages   = {021314},
  year    = {2019},
  doi     = {10.1063/1.5088164}
}

@article{DuaPRXQ2024,
  author  = {Dua, Arpit and Kubica, Aleksander and Jiang, Liang and Flammia, Steven T. and Gullans, Michael J.},
  title   = {Clifford-deformed Surface Codes},
  journal = {PRX Quantum},
  volume  = {5},
  pages   = {010347},
  year    = {2024},
  doi     = {10.1103/PRXQuantum.5.010347}
}

@article{TuckettPRL2018,
  author  = {Tuckett, David K. and Bartlett, Stephen D. and Flammia, Steven T.},
  title   = {Ultrahigh Error Threshold for Surface Codes with Biased Noise},
  journal = {Physical Review Letters},
  volume  = {120},
  pages   = {050505},
  year    = {2018},
  doi     = {10.1103/PhysRevLett.120.050505}
}

@article{tuckett2018tailoring,
  author  = {Tuckett, David K. and Bartlett, Stephen D. and Flammia, Steven T. and Brown, Benjamin J.},
  title   = {Fault-tolerant thresholds for the surface code in excess of 5\% under biased noise},
  journal = {Physical Review Letters},
  volume  = {124},
  pages   = {130501},
  year    = {2020},
  doi     = {10.1103/PhysRevLett.124.130501}
}

@article{BonillaNatComm2021,
  author  = {Bonilla Ataides, J. Pablo and Tuckett, David K. and Bartlett, Stephen D. and Flammia, Steven T. and Brown, Benjamin J.},
  title   = {The XZZX surface code},
  journal = {Nature Communications},
  volume  = {12},
  pages   = {2172},
  year    = {2021},
  doi     = {10.1038/s41467-021-22274-1}
}

@article{TiurevQuantum2023,
  author  = {Tiurev, Konstantin and Derks, Peter-Jan H. S. and Roffe, Joschka and Eisert, Jens and Reiner, Jan-Michael},
  title   = {Correcting non-independent and non-identically distributed errors with surface codes},
  journal = {Quantum},
  volume  = {7},
  pages   = {1123},
  year    = {2023},
  doi     = {10.22331/q-2023-09-26-1123}
}

@article{ClaesNPJQI2023,
  author  = {Claes, Jens and Bourassa, J. Eli and Puri, Shruti},
  title   = {Tailored cluster states with high threshold under biased noise},
  journal = {npj Quantum Information},
  volume  = {9},
  pages   = {82},
  year    = {2023},
  doi     = {10.1038/s41534-023-00677-w}
}

@article{Vargo2014,
  author  = {Bravyi, Sergey and Suchara, Martin and Vargo, Alexander},
  title   = {Efficient algorithms for maximum likelihood decoding in the surface code},
  journal = {Physical Review A},
  volume  = {90},
  pages   = {032326},
  year    = {2014},
  doi     = {10.1103/PhysRevA.90.032326}
}

@article{PoulinQECTN2014,
  author  = {Ferris, Andrew J. and Poulin, David},
  title   = {Tensor Networks and Quantum Error Correction},
  journal = {Physical Review Letters},
  volume  = {113},
  pages   = {030501},
  year    = {2014},
  doi     = {10.1103/PhysRevLett.113.030501}
}

@article{Chubb2D,
  author  = {Chubb, Christopher T.},
  title   = {General tensor network decoding of 2D Pauli codes},
  journal = {arXiv preprint arXiv:2101.04125},
  year    = {2021}
}

@misc{ChubbGithub,
  author  = {Chubb, Christopher T.},
  title   = {Sweep Contractor},
  howpublished = {\url{https://github.com/chubbc/SweepContractor.jl}},
  year    = {2021}
}

@article{TiurevDWCC2024,
  author  = {Tiurev, Konstantin and Pesah, Arthur and Derks, Peter-Jan H. S. and Roffe, Joschka and Eisert, Jens and Kesselring, Markus S. and Reiner, Jan-Michael},
  title   = {Domain Wall Color Code},
  journal = {Physical Review Letters},
  volume  = {133},
  pages   = {110601},
  year    = {2024},
  doi     = {10.1103/PhysRevLett.133.110601}
}

@article{SanMiguelXYZ2023,
  author  = {San Miguel, Jonathan F. and Williamson, Dominic J. and Brown, Benjamin J.},
  title   = {A cellular automaton decoder for a noise-bias tailored color code},
  journal = {Quantum},
  volume  = {7},
  pages   = {940},
  year    = {2023},
  doi     = {10.22331/q-2023-03-09-940}
}

@article{HuangPRXQ2023,
  author  = {Huang, Eric and Pesah, Arthur and Chubb, Christopher T. and Vasmer, Michael and Dua, Arpit},
  title   = {Tailoring Three-Dimensional Topological Codes for Biased Noise},
  journal = {PRX Quantum},
  volume  = {4},
  pages   = {030338},
  year    = {2023},
  doi     = {10.1103/PRXQuantum.4.030338}
}

@article{CamposCompass2024,
  author  = {Campos, J. A. and Brown, Kenneth R.},
  title   = {Clifford-Deformed Compass Codes},
  journal = {arXiv preprint arXiv:2412.03808},
  year    = {2024}
}

@article{RegladeCatQubit2024,
  author  = {R\'{e}glade, U. and others},
  title   = {Quantum control of a cat qubit with bit-flip times exceeding ten seconds},
  journal = {Nature},
  volume  = {629},
  pages   = {778--783},
  year    = {2024},
  doi     = {10.1038/s41586-024-07294-3}
}

@article{Puri2020,
  author  = {Puri, Shruti and St-Jean, Lucas and Gross, Jonathan A. and Grimm, Alexander and Frattini, Nicholas E. and Iyer, Pavithran S. and Krishna, Anirudh and Touzard, Steven and Jiang, Liang and Blais, Alexandre and Flammia, Steven T. and Girvin, Steven M.},
  title   = {Bias-preserving gates with stabilized cat qubits},
  journal = {Science Advances},
  volume  = {6},
  pages   = {eaay5901},
  year    = {2020},
  doi     = {10.1126/sciadv.aay5901}
}

@article{Fowler2012,
  author  = {Fowler, Austin G. and Mariantoni, Matteo and Martinis, John M. and Cleland, Andrew N.},
  title   = {Surface codes: Towards practical large-scale quantum computation},
  journal = {Physical Review A},
  volume  = {86},
  pages   = {032324},
  year    = {2012},
  doi     = {10.1103/PhysRevA.86.032324}
}

@article{Terhal2015,
  author  = {Terhal, Barbara M.},
  title   = {Quantum error correction for quantum memories},
  journal = {Reviews of Modern Physics},
  volume  = {87},
  pages   = {307},
  year    = {2015},
  doi     = {10.1103/RevModPhys.87.307}
}

@article{Klimov2018,
  author  = {Klimov, P. V. and others},
  title   = {Fluctuations of Energy-Relaxation Times in Superconducting Qubits},
  journal = {Physical Review Letters},
  volume  = {121},
  pages   = {090502},
  year    = {2018},
  doi     = {10.1103/PhysRevLett.121.090502}
}

@article{Wang2003,
  author  = {Wang, Chenyang and Harrington, Jim and Preskill, John},
  title   = {Confinement-{Higgs} transition in a disordered gauge theory and the accuracy threshold for quantum memory},
  journal = {Annals of Physics},
  volume  = {303},
  pages   = {31--58},
  year    = {2003},
  doi     = {10.1016/S0003-4916(02)00019-2}
}

@article{Darmawan2017,
  author  = {Darmawan, Andrew S. and Poulin, David},
  title   = {Tensor-Network Simulations of the Surface Code under Realistic Noise},
  journal = {Physical Review Letters},
  volume  = {119},
  pages   = {040502},
  year    = {2017},
  doi     = {10.1103/PhysRevLett.119.040502}
}

@article{RoffeLDPC2023,
  author  = {Roffe, Joschka and Cohen, Lawrence Z. and Quintavalle, Armanda O. and Chandra, Daryus and Campbell, Earl T.},
  title   = {Bias-tailored quantum {LDPC} codes},
  journal = {Quantum},
  volume  = {7},
  pages   = {1005},
  year    = {2023},
  doi     = {10.22331/q-2023-05-15-1005}
}

@phdthesis{KhosravaniThesis2024,
  author  = {Khosravani, Omid},
  title   = {Heterogeneous Quantum Information Processing},
  school  = {Duke University},
  year    = {2024},
  url     = {https://dukespace.lib.duke.edu/entities/publication/49f4294f-e826-4d1f-8b14-1d7324f1b674}
}

@misc{KhosravaniAPS2025,
  author  = {Khosravani, Omid and Escobar-Arrieta, Guillermo and Guti\'errez, Mauricio and Brown, Kenneth R.},
  title   = {Heterogeneous Quantum Error-Correcting Codes},
  howpublished = {Contributed talk, APS Global Physics Summit, Session Q07},
  year    = {2025},
  url     = {https://schedule.aps.org/smt/2025/events/MAR-Q07/13}
}

@article{SteinHetEC2024,
  author  = {Stein, Samuel and Xu, Shifan and Cross, Andrew W. and Yoder, Theodore J. and Javadi-Abhari, Ali and Liu, Chenxu and Liu, Kun and Zhou, Zeyuan and Guinn, Charles and Ding, Yufei and Ding, Yongshan and Li, Ang},
  title   = {{HetEC}: Architectures for Heterogeneous Quantum Error Correction Codes},
  journal = {arXiv preprint arXiv:2411.03202},
  year    = {2024},
  doi     = {10.1145/3676641.3716001}
}

@article{SteinbergHolographic2025,
  author  = {Steinberg, Matthew and Fan, Junyu and Eisert, Jens and Feld, Sebastian and Jahn, Alexander and Cao, Chunjun},
  title   = {Universal fault-tolerant logic with heterogeneous holographic codes},
  journal = {arXiv preprint arXiv:2504.10386},
  year    = {2025}
}

\end{document}